# An Essay on the Relationship between the Mind and the Physical World

Douglas M. Snyder

## Table of Contents

# Preface

I've endeavored in this essay to make my points simply.  If you would like a more technical and detailed discussion of the points that are to follow, you might look at my book, *The Mind and the Physical World: A Psychologist's Exploration of Modern Physical Theory*.  It also is published by Tailor Press.

I would like to thank Dr. Arthur Huffman for thought-provoking discussions and for clarifying various elements of physical theory.  I also wish to thank Donald Musco and Dr. Jeffrey Simon for their incisive editorial comments.

# Chapter 1
## Here We Go!

      We are all used to looking around in the world and considering that we are observing things that are really there, just as they appear, whether or not we are looking at them.  We think that things in the world are not affected by our thinking about them.  We think that the world was there, just as it appears to us now, before we lived and will continue to be there after we are no longer alive.  We know that our bodily actions occur in the physical world, but for the most part we don't think our minds can directly affect the physical world, outside of our intentions, perhaps, that are expressed in our bodies' actions.  Instead, we believe that things in the physical world determine how we experience them.  When we hear something, for example, it is due to vibrations emanating from something in the physical world that travel through the air until they reach our ears.  Contrary to our view of the relationship, or really lack of it, between the mind and the physical world, modern physical theory indicates that the mind directly affects the physical world.  The mind affects both the structure of the physical world and what happens in it.  You probably don't believe me.  But I think you will when we're done.  For now, please bear with me.

# Chapter 2
## Funny Things About Light

Light is very strange, and it is also very important. More important than you think. I know you think it is important because you know that we use light to see. But there is more. Light also is at the root of time, and this is what we need to explore.

Imagine yourself driving down the highway at 50 miles per hour. And imagine further that a car passes you that is going in the same direction as you and that its speed is 65 miles per hour. If I ask you how fast this other car is going relative to your car, your answer will be 15 miles per hour. You would be right. The reason you said the other car is going 15 miles per hour relative to you is because your own velocity and the other car's velocity relative to you need to add up to the velocity of the other car relative to the road.

Now try this. Imagine that this other car is actually a ray of light. Relative to the road, the ray of light has a velocity of $c$ (about 186,000 miles per second). If I ask you what is the velocity of the light ray relative to you, I bet I know what you will say. It is $c$ - 50 miles per hour. You've subtracted your velocity relative to the road from the velocity of the light ray relative to the road. You know what? This is not what happens. If you are not sitting down, I think you should because it is very hard to believe that the velocity of light relative to you is also $c$. I will state this again because it is so surprising. The velocity of light relative to you is the same as it is relative to the road. Why this is the case, no one knows. That this is the case is the basis for Einstein's special theory of relativity. When physicists first realized the velocity of light was the same in situations like I mentioned, they could not believe it. And they did everything they could do to make things add up.

Einstein was a young man when all this was going on, and he had a different idea. He said that we should accept the constant velocity of light in situations like I mentioned. Moreover, he said we should make it the basis for time! Einstein followed through on his plan, and he came to very odd conclusions. They all hinged on one basic result. It has to do with what we call "the same time."

While you're driving down the road, let's say you suddenly see two explosions off to the side of the car, one behind you and one in front of you, at the same time. The explosions in front of you and behind you are the same

# Funny Things

distance from you. I realize this is not an everyday occurrence, but for the sake of argument let's just say it happens. Now if I asked you if a man standing on the side of the road whom you pass in your car just when the explosions occur sees the explosions at the same time, I think I know what you'll say. "Yes, the other man sees the explosions at the same time." But according to special relativity, there is another answer. Einstein decided to define time in situations like the car or the man at the side of the road in terms of the motion of light. Remember that light has the same velocity in all these situations. Because of Einstein's definition of time, we cannot say that the man at the side of the road will also see the explosions at the same time as you driving down the road. Why? Because we no longer have the "adding up of velocities" when the velocity of light is involved. It is the adding up of velocities that keeps the time in your car the same as the time for the man at the side of the road.

Without this adding up of velocities, we have lost the glue for the situations that keeps the observers in them seeing certain things the same. Like how long something is, or how long it takes for something to occur. When we work things out using the motion of light to define time in different situations like I've noted, we find that "the same time" in one situation is not "the same time" in the other situation. We discover that the length of an object along a certain direction in one situation is not the length of the object in the other situation. Also, how long something takes in one situation is not how long it takes when viewed from another situation. So, for example, the man on the side of the road and you in the car come up with different lengths for a stick on the side of road that is facing the direction in which you are traveling. How long it takes for a lizard to crawl from one end of the stick to the other is not the same for the man on the side of the road and you in the car.

I know all of this sounds odd, but this is what experiments tell us is the way things work. Einstein's decision to define time in terms of the motion of light was correct.

Now I want to concentrate on this last thing, how long something takes. We use clocks to determine how long something takes. That's not a surprise to you. Einstein essentially relied on clocks which use the motion of light back and forth to determine how long something takes in situations like our cars and the side of the road that are moving relative to one another. Another very fine physicist, Feynman, pointed out something very interesting. If instead of using light clocks, we used clocks of some other kind, even our ordinary wind-up clocks at home, we would find that they work the same way as the light clocks. Why is this so interesting? The light clocks don't have to be anywhere around the non-light clocks and the non-light clocks work the same way as the light clocks. Remember that light clocks exhibit the constant velocity of light that I discussed earlier, a property that our ordinary clocks don't rely on because their functioning is not based on the motion of light. So why do the non-light clocks work the same way? There is no reason in the physical world why they have to. Feynman said they have to because if they don't, people on the side of the road and in the car won't see the things happening in the world according to the same rules. Now you can imagine what a dilemma this would pose.

Feynman's conclusion is quite unusual. A physicist says that different kinds of things in the world work the same way in our different situations for no other reason than the need for the rules to work the same in them! What is the nature of the rules? The first thing someone thinks is that the rules are part of the physical world itself. But, if this is the case, why don't we see some mechanism in the physical world for the light clocks to affect the non-light clocks? There is no mechanism in the physical world. So the rules are not part of the physical world itself, even though they are the basis for how the physical world works. We must go a step further in trying to discover where the rules are.

Science is concerned with observation. There is no doubt about that. So we have to find some other area open to observation where we can determine the nature of the rules. Now remember, the rule that things have to work the same way on the side of the road or in the car is a fundamental principle of the theory of special relativity.

We need to return to an earlier point. Remember I stated that the length of an object along a certain direction in one situation is not the length of the object in the other situation and that how long something takes in one situation is not how long it takes when viewed from another situation. Which situation is the one where an object is longer and which situation is the one where the object is shorter? Which situation is the one where the duration of how long something takes is longer and which situation is the one where the duration of how long something takes is shorter? That's right. I didn't say. That's because it's both. Both? Yes, both. It depends on in which situation we first define "the same time" and which second. Our situations like the car and the road are moving *relative* to one another. Neither situation can claim to be absolutely at rest.  We saw that this was the case in that the rules for how things work in the physical world need to be the same in our different situations.  It's your choice where you want objects to be shorter and where you want things to take more time.  The situation where you first define "the same time" will find moving objects shorter along a certain direction and require a longer period for things to happen.

Consider a ruler.  A moving ruler is shorter than an identically constructed ruler that is at rest and that is used to measure the length of the "moving" ruler.  In a situation where the "moving" ruler is at rest, this ruler can instead be used to measure the formerly "at rest" ruler which is now "moving."  The ruler that is now "at rest" will find the ruler that is now "moving" to be shorter.  It sounds like a tongue twister because the whole scenario is reciprocal.  We just have to decide which way we want to do our reasoning.  It comes down to: In which situation is "the same time" defined first in our reasoning?

"But," you might say, "an object is an object and a clock is a clock.  Even if things are the way they are described here, we still have something concrete, something the nature of which you can't think away."  True enough it seems, but I can demonstrate that this concrete object or clock doesn't seem so concrete even in a situation where it is at rest.

For this next point, I will rely initially on Einstein's own argument.  Instead of automobiles going down a road, think of a railroad car going down a track.  Lightning flashes hit both ends of the track where the ends of the railroad car are and move toward each other.  For an observer standing on the embankment, the lightning flashes travel equal distances toward each other before they meet.  This observer concludes the lightning flashes occurred at the same time.  But what about an observer at the middle of the railroad car?  If we rely on the time of the person on the embankment, which we are doing, the train moves toward one lightning flash and away from one lightning flash.  You know they won't meet the observer at the middle of the railroad car.  This means these lightning flashes did not strike the points of the track corresponding to the ends of the railroad car at the same time for the observer on the railroad car.

This last statement doesn't sound right because, according to what I stated earlier, light should have the same velocity for the observer on the railroad car that it has for the observer on the side of the embankment.  This

# Chapter 2

issue needs further clarification. At this point though, how do we establish "the same time" in the railroad car? For the light ray that has "slowed" because it moves in the same direction as the train, we can find a light ray to meet it at the middle of the railroad car. But it has to leave the other end of the car where the "fast" one left later than the "fast" one. What does this mean for the length of the railroad car? It means that the length of the car is longer when we define "the same time" for it *after* defining "the same time" for the embankment. If we had gone through this whole sequence of things beginning with "the same time" for the car, it is the distance on the embankment that would have been longer and the length of the railroad car would have been what the distance on the embankment was when "the same time" was first defined for the embankment.

Why did I present this argument now? To point out that the length of a physical object at rest in some situation depends on whether "the same time" is defined first or second when comparing length in this situation to another one moving at a constant velocity relative to it. A similar scenario exists for clocks that are used to measure the durations of occurrences. So you cannot even say, "The concrete nature of a physical object provides something unaffected in its nature by the thought of the person considering the situations."

It is the *argument* on "the same time" in the two situations that is essential to the physical results. Without this argument, the light flashes would meet both observers, the man on the railroad car and the man on the embankment, midpoint in their respective situations (in which they are at rest) because of the constant velocity of light.

Think a little more about these light flashes used to establish "the same time." How could we be dealing with different light flashes for the observer on the embankment and the observer on the train? Each observer can use these flashes to establish "the same time" first in their respective situations before "the same time" is defined in the other situation. Yet when light flashes are used to define "the same time" second in their respective situations, it looks like different light flashes are used than when "the same time" is defined first. I think you can see how the nature of the light flashes depends on the thought of the person thinking about time in the two situations that are moving relative to one another at a constant velocity.

## Funny Things

What about the light and non-light clocks? Science and observation. What else can be observed if not the physical world itself? Our thought processes. These are also involved in the nature of time in special relativity: 1) as an observer in one of our situations, and 2) as a theoretician working through the arguments in the special theory, in particular the determination of "the same time" in our situations. There will be more support for this conclusion shortly.

Chapter 3

A Slight Digression

It's not easy figuring out the order in which to present the topics in this essay. In fact, knowing how to make logical arguments is never really easy. It requires the ability to make sense of things, to find order in them. People vary in their ability to find order in things, including those in the physical world. This ability to order things is distributed in a manner that resembles a curve that will be discussed in a later chapter on statistical mechanics. It is called the normal curve. In science, this order that people find in the physical world is anchored in theories, logical arguments, made about the world that are then subject to experimental test.

The arguments I have presented so far, and will continue to present, are concerned with the thinker thinking about the physical world or the observer observing the physical world. They are not concerned with the physical world as if it were a peach pie sitting right in front of your eyes waiting to be eaten. They are about the ordering process itself. That's why the points I am making have not been considered very much before. Physicists, though experts on the physical world, are not trained today to think about the process of thinking. They are trained to investigate phenomena in the physical world that are kind of like our peach pie. They probably don't want to devour most of the things they study, but generally they are objects for them. Really, they are objects for their thought and perception. The reason I can make the points in this essay is that I was trained differently. I was trained as a psychologist and thus to think about the thinker who is thinking about things. And after this training, I became interested in physical theory. The resulting perspective is one where I look at the physical world in terms of the theory constructed by physicists but, in addition, I look at the theorizing itself and the observational activities of physicists and others watching the physical world.

Chapter 4

Conservation Laws and What?

You've heard of conservation laws, how certain things don't change in a group of physical objects that are interacting with each other if nothing from the outside disturbs them. The most well-known is the conservation of energy. It says that the energy of this group of objects doesn't change. For example, consider hanging two polished steel balls on a string from a piece of wood and pull one back and then let it go so it hits the other ball square in the middle. The other ball will move outward and then swing back and hit the ball that first hit it. Now this can go on for a while, especially if there is no air present around the balls. What's going on? Energy is being conserved. That's why the balls keep hitting each other.

The conservation of energy has a very important feature. It depends on physical law not changing over time. One might assume that the laws governing the physical world could change as time elapses, but the fact that they don't explains why the balls continue to hit each other for a very long time.

Is there anything else to consider? There is. Measuring time requires a scale, just like measuring space requires a reference frame. What is meant by a reference frame? So far we've discussed rulers, distance, the length of an object. Actually, what we have been discussing is a reference frame. It is just a coordinate system that allows you to measure the position of things in space. In your automobile, you have such a coordinate system and the man at the side of the road has such a coordinate system. Just like you move relative to the man on the side of the road, so your reference frame moves as well. A reference frame needs something physical associated with it, but it is up to us to determine what that something is.

To show that a reference frame really doesn't depend on any physical object external to you, imagine that a rock is at rest relative to the man on the side of the road and the rock is considered the physical anchor to which the man's frame of reference is attached. The reference frame can be considered an imagined set of giant antennae stretching up-down, in-out, and side-to-side, the antennae in each direction divided into pieces of the same length. Now imagine that the rock is set in motion. Maybe someone comes along, roles it over a few times, and the rock keeps going. If the man wants to measure the distance of things from each other, what does he do? He would do the same

Chapter 4

thing he did before the rock started rolling.  He needs to pick a new rock, or some other object, to attach his giant mental antennae to.  That's all.  To put it differently, a particular rock has nothing to do in an essential way with his reference frame.  Rather, the reference frame depends on him.  If the reference frame was not at rest relative to the observer, he would not know what's what concerning the motion of objects.

The observer is at rest in his reference frame.  That's just another way of saying that the observer's reference frame is at rest relative to him.  That the observer is at rest in his reference frame is borne out by experience, and we will come back to this point in a minute.  There's another factor to be considered that supports a connection between a reference frame and an observer.  The relativity principles in physics maintain that there is no preferred reference frame that is absolutely at rest.  In Newtonian mechanics, for example, physical phenomena are described the same by observers in either of two situations moving at constant velocity relative to one another in a straight line (like the railroad car and the embankment).  It shouldn't make a difference for physical description in our situations in special relativity as well.  (This is not to say that where "the same time" is first defined is unimportant.  Rather, once the situations are defined and used in the arguments concerning length and time in the same way, for example, physical description in these situations should be equivalent.)

You may ask me what I mean when I say an observer is at rest in a reference frame?  How can an observer riding in an automobile feel himself to be at rest?  There is no problem with the observer on the side of the road because anyone standing on the side of a road feels like he is at rest.  But wait, isn't this man also moving?  Isn't the observer on the side of the road moving with the earth as it rotates on its axis and as it moves around the sun?  Actually, the earth is moving pretty fast.  Surely, if a person were in a car moving as fast as the earth is moving, he would feel it.  What is going on here? Now think.  When you are riding in a car and you close your eyes, what do you experience?  You feel a certain vibration in your body.  Since you've ridden in a car before, you conclude that this bumpiness means you're riding along a road.  If you hadn't ridden in a car before, you just wouldn't know what the bumpy feeling meant. Let's say the bumpiness is completely absent because you're riding on a very smooth road in a luxury car.  If the car was moving in a straight line, you would feel that you were at rest.  That's right, at rest.  Just think of yourself in an airplane when you're in the air.  You don't

Conservation Laws

feel like you're moving. In fact, it may amaze you when you disembark from your first flight in a place distant from where you began that you actually got there without more of a feeling of motion. Imagine how the astronauts felt on their trip to the moon.

What is responsible for the observer's being at rest in his reference frame? A physical cause cannot be identified. We choose a physical object to attach our reference frame to that is at rest relative to us. A psychological phenomenon is at work here, but that shouldn't be too surprising because what's a reference frame anyway? It's really a cognitive construction that we use in living in and studying the physical world, such as saying the distance between two points is one foot. There is nothing that says that a particular mental construction has to be used in constructing a reference frame. I could have said the distance between the two points is two feet and no one could have told me it's not. This point is so basic to our description of the physical world that any other results indicating that there are different ways to describe the motion of physical objects should not be too surprising. These other results depend on the flexibility in constructing a reference frame.

Just like a reference frame, the time scale we use needs to be associated with something in the physical world. We can determine what that something is. It could be a clock with minute, hour, and second hands, or it could be some other physical process in the world. Even a person's getting older could be the basis for a time scale. Now we can return to physical law remaining the same over time. In testing whether physical law doesn't change over time, you probably thought the only kind of test would be to let the clock tick to see if this is the case. If you think about it though, you could just shift your time scale forward or backward any way you wish. There really isn't any difference in watching our clock or shifting the time scale mentally. Physical law wouldn't change. This may seem silly, but I assure you it is not. It's the simplest and most elementary things we need to be careful about.

Perhaps it will seem a bit less silly if we consider space. We have already observed that changing the time of an occurrence doesn't change physical law. A similar thing happens in space. Let's say we change the position of an object from one point to another point along a straight line. We see that physical law doesn't change. Besides moving the physical object itself in space, we can change the position of the ruler itself, for example, that acts as our straight line in the physical world. We can also shift the straight line mentally in relation to the physical object.

Chapter 4

      We also could rotate the position of an object in space from one point to another, and we would find that physical law is no different at either of these points.  This rotation can be accomplished, for example, by rotating the object itself or by mentally rotating the axes while leaving the object untouched.  So you see certain conservation laws (those associated with physical law not changing over time, with the change in an object's position along a straight line, and the rotation of an object from one position to another) are associated with our thinking.  Certain physical and mental actions are equivalent with regard to the conservation of particular physical quantities.  For the record, these quantities are energy, linear momentum, and angular momentum.

Chapter 5

Is Gravity a Problem?

It is surprising Einstein himself didn't say that the mind is linked to the physical world. He surely saw what Feynman saw, namely that the need to maintain the integrity of the special theory was the basis for various physical results. He even began discussing "the same time" in his first paper on special relativity in a non-quantitative psychological way. Einstein, though, didn't think to make this connection, even though he knew that in general relativity and quantum theory a mental tie to the physical world is also plausible, even more plausible than in special relativity. Let's take a look at general relativity.

Einstein thought there was no reason the laws of physics should not hold in all reference frames. I have to confess something. The reference frames with regard to special relativity presented earlier have included factors like gravity and friction. Really, with regard to special relativity, it's best to imagine all these previous situations occurring deep in outer space far away from any large bodies. It's strange to think of automobiles and railroad cars in outer space, so instead think of space ships. In outer space far away from any large bodies, there is no gravitational influence on the space ships and there is no friction. There, when something moves in a straight line and with a certain velocity, it just keeps moving in a straight line with that velocity unless an external force impacts it. When something is stopped, it remains stopped in the absence of an external force. When you push something, the object accelerates in the direction of the force exerted on it in a way that depends on the mass of the object. Also, when you push something, it pushes back. These ways in which objects behave are called Newton's laws of motion.

Special relativity encompasses these laws and goes beyond them. General relativity encompasses special relativity and goes beyond it. We need to back up, as I've begun to do, and think about Newton's laws of motion a bit more. This will allow us to see how Einstein developed general relativity and show us how the mind is linked to the physical world in this theory.

After developing special relativity, Einstein said the laws of physics should account for gravity and apply to accelerating reference frames as well as to reference frames deep in space far from large objects where Newton's laws hold. Why could he say this? Well, it was known that accelerated

Chapter 5

motion is the same in these reference frames deep in space when they are moving relative to one another with constant speed and direction. But this fact about accelerated motion works only for these reference frames deep in space. So the supposedly "absolute" nature of accelerated motion *depends* on a *certain kind* of reference frame. The other thing that appeared absolute in a different sense were Newton's laws of motion. It could be shown that if they held in one reference frame deep in space, they would hold in all other reference frames deep in space and far from large objects which had a constant speed and direction relative to the original reference frame. Newton's laws of motion would not hold, though, in accelerating reference frames. So the apparently inviolable character of physical law also really depended on a certain kind of reference frame.

Now Einstein, smart as he was, had some unusual thoughts about the acceleration of an object in one of these reference frames deep in space. He deduced that the acceleration feels the same as gravity for an observer in an accelerating reference frame deep in space. He also figured out that an observer accelerating due to gravity would think he was in a reference frame like one deep in outer space. I should tell you about these situations.

Let's start with the second one first. Think of a man in an elevator in a skyscraper. Suddenly, the elevator cable snaps and the elevator falls toward the ground. The man experiences the physical world as if he is in outer space. For this man, objects in the elevator act in accordance with Newton's laws of motion and do not appear to be under the influence of gravity. If he lets go of a ball without pushing or pulling it, for example, it will stay even with him, just as it would in outer space. Let's see how Einstein (1938/1966) put it. "If the observer pushes a body in any direction, up or down for instance, it always moves uniformly so long as it does not collide with the ceiling or the floor of the elevator. Briefly speaking, the laws of classical mechanics are valid for the observer inside the elevator [who does not experience gravity]" (p. 215).

Let's consider the first situation of a man in a windowless room in deep space. The room is towed by a rope attached to one end so that the room accelerates at a uniform rate. What does the man inside the room experience? He feels as if he is in a gravitational field. For example, if the man in the room hangs a rope from the ceiling and attaches an object to the bottom, the rope hangs toward the floor. What does the man in the room think? He thinks, "The suspended body experiences a downward force in the gravitational field, and this is neutralised by the tension of the rope" (Einstein,

# Gravity

1917/1961, p. 68). What does an observer outside the room deep in space think? He thinks, "The rope must perforce take part in the accelerated motion of the chest [the room], and it transmits this motion to the body" (Einstein, 1917/1961, p. 68).

Einstein saw that an accelerating reference frame, the accelerating room, was equivalent to one of these reference frames deep in outer space that obeyed Newton's laws of motion, but with a slight twist. The slight twist is that such a frame would experience a gravitational field. This is really not a problem because a gravitational field exerts a force on objects and this force affects objects the same way that I indicated objects are affected by forces deep in space, namely in accordance with Newton's laws of motion. So Einstein could seriously consider that the laws of physics should hold for all reference frames. That these different reference frames are capable of describing the same physical phenomenon is the heart of the principle of equivalence in general relativity.

Also, Einstein had a way to develop a system to measure space and time, really space-time, in accelerating reference frames and reference frames experiencing gravitational fields. What Einstein did was to break up a uniformly accelerating reference frame into a sequence of very tiny reference frames. Then each tiny reference frame is like a reference frame in special relativity. Two of these tiny reference frames move at a constant velocity relative to one another. We've seen how we can develop the space-time relationships between two inertial reference frames moving at a constant velocity relative to one another. Einstein could develop a space-time metric by considering these tiny reference frames in relation to one another. Some other well-known physicists commented on the importance of special relativity to general relativity. They wrote:

> General relativity is built on special relativity. (Misner, Thorne, & Wheeler, 1973, p. 164)

Elaborating on this statement, they said:

> A tourist in a powered interplanetary rocket feels "gravity." Can a physicist by local effects convince him that this "gravity" is bogus? Never, says Einstein's principle of the local [over a small area] equivalence of gravity and accelerations. But then the physicist will make no errors if he *deludes* [italics added] himself into treating true gravity as a
- 15 -

Chapter 5

> local illusion caused by acceleration. Under this delusion, he barges ahead and solves gravitational problems by using special relativity: if he is clever enough to divide every problem into a network of local questions, each solvable under such a delusion, then he can work out all influences of any gravitational field. Only three basic principles are invoked: special relativity physics, the equivalence principle, and the local nature of physics. They are simple and clear. To apply them, however, imposes a double task: (1) take space-time apart into locally flat pieces (where the principles [of the special theory] are valid), and (2) put these pieces together again into a comprehensible picture. To undertake this dissection and reconstitution, to see curved dynamic space-time inescapably take form, and to see the consequences for physics: that is general relativity. (p. 164)

Einstein proceeded to develop physical law so that it would also encompass accelerating reference frames and reference frames experiencing gravitational fields. He also knew that he could provide a physical basis for an object's gravitational mass being essentially equal to the resistance of this object to acceleration by a force. This equality is important for everyone, especially Galileo who found that the acceleration of objects in the earth's gravitational field was the same, regardless of the type of object or its gravitational mass. That the man in the room accelerating in deep space could describe the physical world as well as someone outside watching the room accelerating requires that this equality hold. The equivalence of their descriptions forms the foundation for the general principle of relativity and the across-the-board application of the laws of physics in different types of reference frames.

Here's the punch line, which you may have already deduced. The difference in the description of the physical world between the accelerating observer in deep space and an observer who is watching the room the man is in accelerate is due to both observers being at rest in different reference frames. Remember a reference frame is a cognitive construction that an individual employs in observing the physical world. We have seen in special relativity the importance of an observer's being at rest in a certain kind of reference frame. Einstein extended this idea of an observer's being at rest to accelerating reference frames as well. The man in the falling elevator and the

# Gravity

man in the spaceship feel themselves fundamentally to be at rest. It's all the "extras" that make people think they are moving. Einstein (1917/1961) discussed a passenger on a train where the brakes are suddenly applied and the passenger experiences a powerful "jerk forwards" (p. 62). Einstein said:

> It is certainly true that the observer in the railway carriage experiences a jerk forwards as a result of the application of the brake, and that he recognises in this the non-uniformity of motion (retardation) of the carriage. But he is compelled by nobody to refer this jerk to a "real" acceleration (retardation) of the carriage. He might also interpret his experience thus: "My body of reference (the carriage) remains permanently at rest. With reference to it, however, there exists (during the period of application of the brakes) a gravitational field which is directed forwards and which is variable with respect to time. Under the influence of this field, the embankment together with the earth moves non-uniformly in such a manner that their original velocity in the backwards direction is continuously reduced (pp. 69-70).

Really, the passenger in the railway car has to see himself at rest or else we will again be faced with privileged reference frames. The observer's being at rest in his reference frame in the special theory of relativity is psychological in nature. His being at rest in his reference frame in the general theory of relativity is psychological in nature too. Remember that the general theory relies on the special theory for measuring lengths and durations. So the psychological parts of the special theory, notably how we decide on "the same time" in reference frames moving at a constant velocity relative to one another, hold for the general theory as well.

A final note: Gravitational mass itself, terra firma, may be simply a matter of reference frames. Maybe this shouldn't be so shocking after Einstein showed that mass is really equivalent to energy, and it is energy that is conserved, not mass as we usually think about it. But that's a lot different than saying "reference frames" because reference frames clearly depend on the mind.

# Chapter 6
# Quantum Mechanics – No, It Can't Be!

Imagine that you are seated in front of a big screen. On the screen, shapes gradually take form, dot by dot, kind of like a painting by Seurat. If you watch the screen over a long period of time, you see a form take shape. It looks like a range of hills where all the hills are skinny and separated by deep valleys. In the middle of the screen is the highest hill and as you move to either side, the hills gradually become smaller. The left and right sides of the hills on the screen look the same.

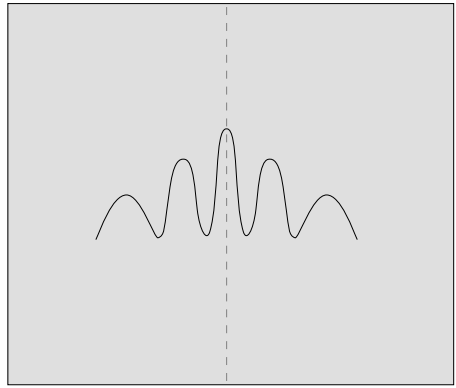

Now the other picture that you see takes shape dot by dot is just one wide hill, perfectly balanced between left and right of the middle. It looks like this.

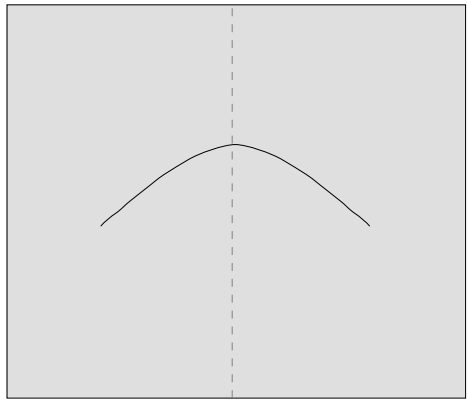

# Quantum Mechanics

To be honest with you, these two pictures are not actually what an observer would see.  But they portray accurately essential elements of the physical situation that an observer would indeed see.  What the vertical positions of the hills represent in reality is this: The higher the location on a hill, the more dots appear at that horizonal location in the picture.

You notice a thin wall.  This wall and our screen line up perfectly, the wall in front of the screen.  The wall has two slits in it, each one a little to one side of the middle.  Now there's also a light just behind the wall and smack in the middle between the two holes.  Looking down, one sees the wall, the screen, and the light line up this way.

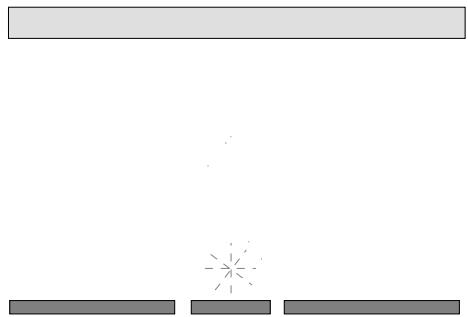

You notice the following.  If the light is off for a while, we see the skinny hills.  If the light is on for a while, we see one wide hill.  In each case, we start with a blank screen.  Now since you are a very careful investigator, you notice another thing that may be important.  When the light is on, there are sudden light flashes to the left and right of the light that are associated with the dots appearing on the screen.  The flashes to the left happen by the left slit and the flashes to the right happen by the right slit.  Each flash to the left or right of the light is associated with a subsequent dot appearing on the screen.  The flashes occur one at a time, and so the flashes and subsequent dots on the screen can be tracked.

The question naturally is, what is responsible for the different pictures that appear on the screen, point by point?  You probably think it's the light being turned on or off, or more specifically the light flashes that occur only when the light is on.  That's a pretty good guess, one that most physicists have made.  But think about the following experiment.  Move the light to one side, say the left side, so that you only get light flashes from one slit in the wall.

Chapter 6

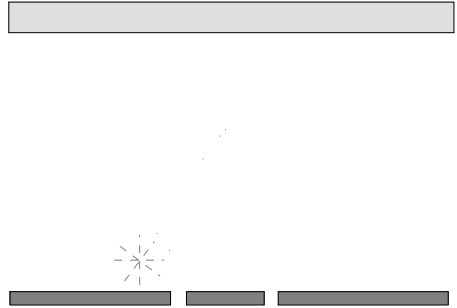

Now which pattern do you think you'll get?  Maybe you'll say a mixture of skinny hills and one wide hill.  What you get is one perfectly balanced wide hill *just like* the one that occurred when the light was in the middle between both slits.  So then, like any good investigator, you ask yourself what can cause the one wide hill pattern that is not present when we get the many skinny hills pattern.  Is it the light flashes at both slits? No. It's the light flashes at one slit?  Well, track the dots that are *not* associated with light flashes, the dots that occur when there are no flashes.  They form a certain pattern, another kind of wide hill that is skewed to the right.

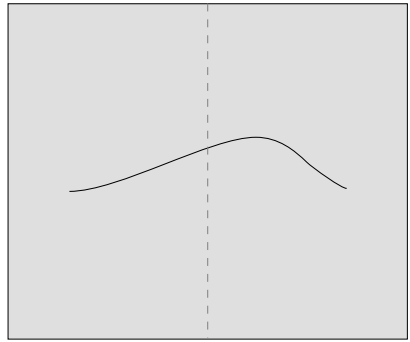

You see that there are dots at places where when the light is turned off there aren't many dots at all.  (Remember when the light is turned off, we get the many skinny hills pattern where there are some deep valleys.)  This pattern is just like the pattern we would get if we placed the light in the middle between the two slits and looked at the pattern formed by the dots associated with the flashes of light at the right slit.  What is going on?

# Quantum Mechanics

Something is happening at the slit away from the light where there are no light flashes. It can't be the light flashes at the slit where the light is. What else? That's a good question. First let me give you one other piece of information.

Take a scenario where the light is located near the left slit and you know how long a specific flash has to take place. If it takes place in that time, okay. If not, it will not take place at all. Let's say the light is turned off before the time has elapsed for a specific flash, even just before the time is up, and a flash has not occurred. What do you think happens? We get the skinny hills pattern if this is repeated many times. Why? The reason is our *knowledge* has not changed. We haven't seen a flash, and we haven't allowed the full time for the possibility of a flash, so we are unable to deduce any associations between specific light flashes at the wall with the two slits, or their absence, and specific dots at the screen that subsequently appear.

Knowledge is what's behind our seeing one wide hill when the light is moved near one slit. In this scenario, there are no light flashes near one of the slits. The fact that there are light flashes at the other slit that can be associated with dots on the screen indicates that the other dots on the screen can be associated with there *not* being light flashes at the other, unilluminated slit. Alternatively, if there were a light near the unilluminated slit, there would be light flashes at this slit that could be associated with these other dots. By extension, knowledge is also behind our seeing one wide hill when the light is in the middle between both slits and we get light flashes near both. We have determined the common factor in the different scenarios associated with the change from the skinny hills to the one wide hill: knowledge.

Also, the shape of the pattern of dots associated with no light flashes in the scenario where the light is located near one slit is similar to the shape of the pattern of dots associated with light flashes from what was the unilluminated slit when the light source is between both slits and is on all the time. That these shapes are similar supports the thesis that physical interaction is not responsible for a specific pattern of dots at the screen that is associated with light flashes near slits in the wall. Further, when we get the one perfectly balanced wide hill pattern, it is the summation of two skewed wide hill patterns, one skewed to the right and one to the left. The wide hill pattern skewed to the right is the pattern formed by the dots associated with the flashes of light at the right slit. The wide hill pattern skewed to the left is the pattern formed by the dots associated with the flashes of light at the left

Chapter 6

slit.  We also get this perfectly balanced wide hill pattern when only one of the slits is illuminated.

These light flashes and dots could involve a variety of small particles, but let's say for example they are electrons.  A light flash by a slit in the wall occurs when an electron passes through a particular slit and is struck by light from the light source, and a dot occurs when the electron strikes our screen.  So you could say we see one wide hill when we know through which slit electrons pass.  In the absence of this knowledge, we see many skinny hills.

On to something else.  I should tell you briefly about Schrödinger's cat.  You might have heard about the cat that is neither alive nor dead in a certain situation, but is kind of both until an observation of it is made.  This situation involves a measurement of a tiny particle where the state of this particle may change.  The state of the cat is tied to the state of the particle.  Many maintain that which state characterizes the cat is tied to the observer's recognition that the cat is either alive or dead, no matter how far away the cat is from the observer.  This means even across the universe.  Even then, with the observer's observation, the state of the cat changes immediately, and it is either alive or dead.  That's it.  The straightforward character of Schrödinger's cat does not make it less remarkable.  It makes it more remarkable because of its basic nature.

I hope all of this information has not upset you.  You know, you look a little red.  Maybe you should take your temperature to be sure you don't have a fever.  But, uh oh, as I will explain, temperature is tied to the mind.  Can a person rely on anything anymore?

Chapter 7

Oh No!  Temperature Too?

In order to demonstrate a link between temperature and the mind, we first have to discuss some concepts from statistics.  Consider 10 pennies that are equally weighted and flipped in a way that does not favor either heads up or tails up.  Also, allow that each coin flip does not affect any other coin flip.  In this situation there is no reason why a head should come up rather than a tail.  So the most logical thing to decide is that head up or tail up for each penny is equally likely.  If we want to figure out how many ways the pennies can land heads up and heads down, our task is relatively easy.  For example, for the combination of one head up and 9 tails up, there are 10 ways the pennies can achieve this combination.  One arrangement of one head up and nine tails up is:

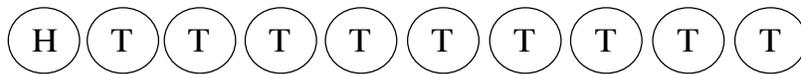

Each of the 10 ways is equally likely to occur when the pennies are flipped.

Now consider five heads up and five tails up.  There are many more ways that the pennies can achieve this combination than they can for one head up and nine tails up.  Just think that instead of shifting the head up penny over one place at a time, there are five heads up pennies that can be shifted throughout the 10 pennies.  In fact, the equal number of heads up pennies and tails up pennies provides the greatest number of ways the pennies can be arranged of the different possible combinations of heads up and tails up.  This last point holds for any even number of pennies.

The next drawing is a graph showing the distribution of heads up pennies when we are dealing with 100 pennies instead of 10 pennies.  The left edge shows the relative number of ways 0 pennies with heads up can be obtained.  The right edge shows the relative number of ways 100 pennies with heads up can be obtained.  The middle shows the relative number of ways 50 pennies with heads up can be obtained.  This graph yields a curve that looks like something called the normal curve.  Since all sequences of pennies are equally likely, this graph also tells you the likelihood that the pennies will be in a particular combination of heads up and tails up.

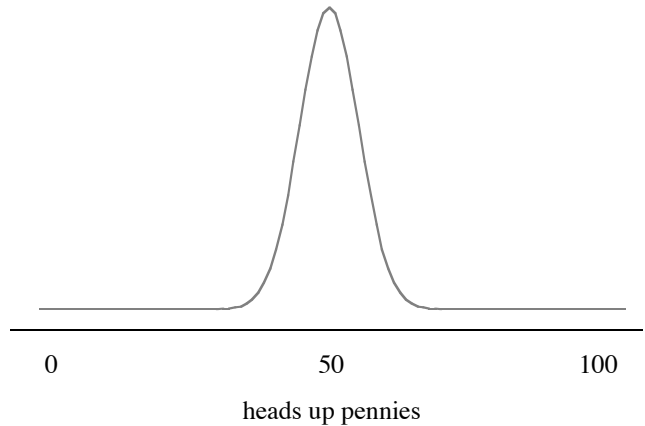

0     50     100
heads up pennies

Now imagine that we have a very large even number of pennies in one group. And imagine that we have another group of pennies, say 10, in another group. These groups are in separate containers, like so.

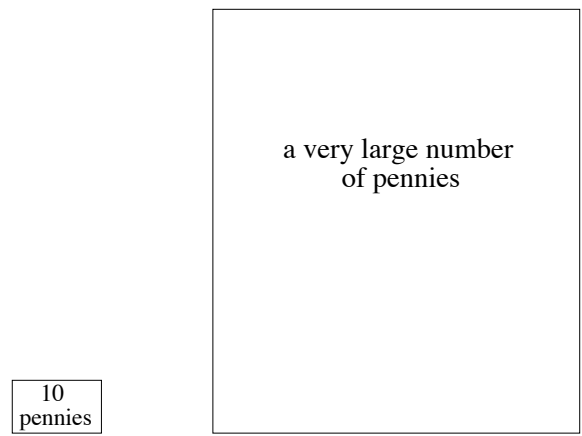

Each of the pennies in both of the containers is always able to flip from heads up to tails up or from tails up to heads up. When in the air, pennies from one group cannot fall into the other group. Imagine that the pennies in the containers are separated from all other physical phenomena and that the pennies in each container can change from sequence to sequence through flipping. The pennies are not stuck in one sequence. If no other conditions are placed on the groups of pennies, what is the most likely combination of heads up and tails up that will be obtained at any time when both groups of pennies are considered? It is 5 heads up and 5 tails up in the

# Oh No!

group of 10 pennies and 1/2 of the pennies in the other group heads up and 1/2 of the pennies tails up.

Now let's add the following condition. Say that a certain amount of energy is associated with the pennies in one container and another amount of energy is associated with the pennies in the other container. Pretend there is a device outside the container that produces an energy field that affects the pennies in the group. Allow that a penny with head up is associated with a different amount of energy than a penny with tail up. Take the container with 10 pennies for example. Instead, of getting 5 pennies with heads up and 5 pennies with tails up as the most likely outcome, we have to take into account the total energy of the 10 pennies.

Each combination of heads up and tails up for the 10 pennies will have a different total energy associated with it. Let's say for example that the total energy of the 10 pennies is such that we much have 3 heads up and 7 tails up. Then it does not matter which of the pennies are heads up and which tails up. In fact, imagine for example the pennies are wired to each other and that the wires have no other effect on the pennies other than to allow them to exchange energy with one another so that they can flip from head up to tail up or from tail up to head up and keep the energy of the pennies in their group the same. Just the combination of heads up and tails up have to remain the same to keep the energy of the 10 pennies the same. The same reasoning holds for the very large number of pennies in the other container.

Instead of being physically separated, let the containers with the pennies touch each other. The walls of the containers can now exchange energy, and this energy is shared among the pennies.

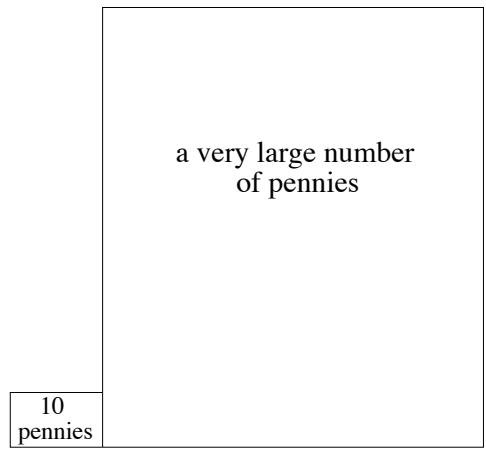

# Chapter 7

Both groups of pennies share a certain fixed amount of energy which is the sum of the energies of the two groups of pennies considered separately. As noted, this energy can be exchanged between the two groups of pennies through the walls of the containers, but the total amount of energy does not change. Then in each group of pennies different combinations of head up and tail up pennies are associated with different amounts of energy, but the total energy of both groups remains constant.

Now we have the interesting question of what is the most likely combinations of heads up and tails up in each of the two groups of pennies? The answer is that the energy is in large part distributed so that the combinations of heads up and tails up pennies for the two groups yield the highest number of ways the pennies can be arranged in different sequences, given the constraint that the energy of the two systems together is constant. Why is this the case? Only one reason. Each sequence of heads up pennies and tails up pennies in a group is equally likely when a group of pennies is considered in isolation, like we did at the beginning of the chapter. When the containers are brought into contact with each other, a new energy constraint has been introduced, the total energy of the pennies in the two containers. So now the pennies have "room" to change their combinations in each of the containers so long as the total energy remains the same. Without any other physical restriction or any physical law governing how the pennies need to be arranged in their respective containers, it makes perfect sense that the two groups of pennies would tend toward the combination for each group that would maximize the number of ways the pennies in both groups could be arranged.

About now, you might ask yourself why is he going to all this trouble to discuss these things? First, this scenario is behind the equilibrium of temperature that two physical systems tend toward when they are brought into the kind of contact I just discussed. A good approximation of this is when you put a thermometer in your mouth to take your temperature. The thermometer tends toward the same temperature as you. The thermometer is like the small system of pennies, and you are like the large system of pennies. Compared to you, the thermometer has a very limited number of possible sequences of its parts, and so the overall number of sequences for you and the thermometer is largely determined by you.

As you can see, physical systems in statistical mechanics are made up of things like the atoms or molecules that comprise you and the thermometer,

# Oh No!

for example.  The pennies, though, are a good way to discuss the statistical concepts.  Instead of heads up and tails up, we would be concerned with velocities, for example, of atoms and molecules.

Consider the following problem.  In the physical world, we only get one sequence of pennies at a time.  Correct?  Hard to argue with that one.  But then how do we get results concerning things like temperature that involve lots of sequences, specifically all the possible sequences for all the possible combinations?  Physicists call this last group a representative ensemble of systems like the actual physical system.  "But," you might ask, "where did this concern with *all* of the sequences come from?"

Certain physicists said that we can consider that the sequences occur quickly over time, one at a time, and the length of a measurement of temperature, for example, takes into account the passage of these systems through various sequences.  In fact, that is what physicists first thought and that seems at first to be the most natural explanation.  A physicist by the name of Tolman made it very clear, though, that this attempted solution does not work, and it is accepted today that the different sequences need not occur one at a time.  This attempted solution doesn't work because we are concerned with the application of statistical methods to physical processes, and it is the integrity of the statistical methods that is of primary concern.  When the different sequences occur is not of great significance.  The important thing is that they are all available due to the flexibility in the physical systems involved, and thus they all contribute to various quantities of the physical systems like temperature.

Oh, this isn't supposed to be possible because only one sequence can exist at a time.  Another physicist, Kittel, wrote that the different sequences exist at one time and that the representative ensemble of sequences is an *intellectual* construction.  What does this mean?  It means that people *think* it.  "What," you say, "people think it?"  Yes, that's right.  The physical world indicates that their thinking is part of what lies behind the temperature of a physical system, including your body.  It also lies behind the conduction of nerve impulses in your body as well the operation of the battery in your car.

Remember the equal likelihood of the different sequences.  Why did I focus on this?  Wouldn't it have been easier to state some law about how the pennies would land?  Maybe they were flipped a certain way that resulted in their landing in a certain sequence.  Then physical law could tell us why the

Chapter 7

pennies land the way they do. This is an interesting argument, and when you look at each individual penny, one can see how it lawfully flies in the air. When we consider the pennies as a group, though, we are dealing with statistics, not law. (I have to admit, it really is ideal to think of individual pennies with a perfect weight distribution and perfectly flipped, but it really is statistics that we are dealing with.) What I mean is that we are dealing with the *possible ways* the pennies can fall into a sequence. This is a lot different than how individual pennies fly. Knowing how each penny flies doesn't help a lot because one would have to know how the pennies had such widely different initial conditions when they were flipped. At some point, a statistical assumption must be made. Physical law only deals with the pennies once we know their initial conditions.

Why make the assumption that the various possible sequences of pennies with heads up and pennies with tails up are equally likely? Now, this is important. It is the only reasonable assumption that can be made if physical law does not apply. I would like to take credit for this last statement, but I can't. Remember Tolman, the physicist. He made it.

Again, we're back to a mental-sounding concept: assumption. We're back to thinking about the physical world that leads to correct predictions about measurements, measurements for which there is no reasonable physical explanation. We come to this conclusion because physical law cannot be the basis for these predictions.

You may have noticed that the discussion in this last section has really been about probability. What the physical world has furnished us is a situation to let probability work. The probabilistic considerations that have been discussed could have involved non-physical things just as well as physical things. It wouldn't have made any difference regarding the predictions. For example, it may interest you that intelligence as measured by IQ tests is distributed in about the same way as our group of 100 pennies. That's right. It is approximately normally distributed. I mentioned this earlier when I wrote about finding order in the world. So probabilistic considerations apply to IQ test results as well. And you know what probabilities are about. They are about knowledge of what will happen.

Maybe I should have told you at the beginning of this chapter on statistical mechanics that probabilities are really concerned with *knowledge* about what will happen and that statistical mechanics is really about

# Oh No!

probabilities. But I thought you wouldn't really take it very seriously. Now I think you do.

To show you how good the results of statistical mechanics are, here is a quote from Einstein on thermodynamics. Statistical mechanics provides a deep theoretical explanation of thermodynamics.

> It [classical thermodynamics] is the only physical theory of universal content concerning which I am convinced that, within the framework of applicability of its basic concepts, it will never be overthrown. (Einstein, 1949/1969, p. 33)

It may well be that of all the areas of physical theory statistical mechanics shows most clearly that non-physical processes are involved in the functioning of the physical world.

Chapter 8

Oh, I've Just Fallen Down, or Is That Up?

Now, if your head is spinning, I really can't blame you. The involvement and affect of mind in and on the physical world are so basic that they are a bit discomforting. In fact if you felt that up is down and down is up, I would understand. But I would understand for an additional reason besides your feeling a bit unbalanced at the moment. Maybe what's coming will seem like bad news to you. Your feeling that up is down and down is up is not far from the way the world is. This is where the discipline of psychology really enters the picture. Research from psychology indicates that what we think is up and down in our visual experience and in the physical world is dependent on various factors pertaining to the observer. It seems that most important are the coincidence of visual and tactual (touch) sensations. Other factors involve kinesthetic stimuli (our own sensations of our muscles) as well as gravity. But these latter factors seem to be less significant. Thus, if the coincidence of visual and tactual sensations is rearranged from what normally occurs, we again get a sense of up and down in visual experience that feels as natural as the first did and which the observer does not readily distinguish from the first unless prompted to do so. Competency in visually guided action returns.

I should explain this research a bit more. At the very end of the 1800s, a psychologist by the name of Stratton wore an apparatus on his head that rotated incoming light to one of his eyes 180º. The other eye was covered. On the days he wore this contraption and went about many of his normal activities, he progressively adapted to inversion of the incoming light along the lines of the results I mentioned. Other psychologists using different optical devices found similar results. In the case where a psychologist thought his own findings did not support these results, many of his results indeed were in accord with them. A very prominent psychologist by the name of Boring thought they were too.

Let's take a look at one implication of this research. Consider an observer A, who isn't wearing any optical contraption and who looks at an arrow in the world that is pointing up. The retinal image of the arrow, the pattern of light on the retina that forms an arrow, is inverted. It points down. That's how the eye usually works.

# Oh, I've Just Fallen

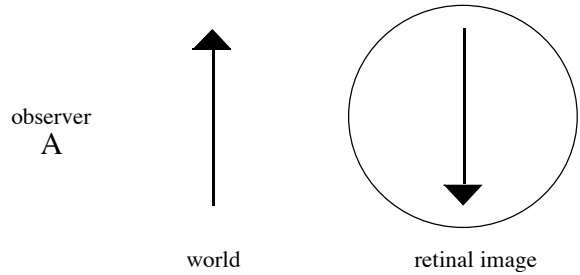

Consider another observer, B, who wears an apparatus like Stratton's and whose vision has adapted. If he looks at the arrow in the world that is pointing up for observer A, the retinal image of the arrow points up and observer B also sees the arrow pointing up.

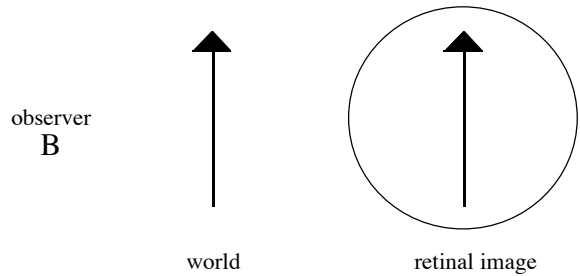

Now this same retinal image for B can also occur for an observer like A (call him A′) who does not wear an optical apparatus. When it does, the retinal image is associated with an arrow in the world pointing down for A′. Thus, the arrow in the world for B points in the opposite direction to that which would be found by A′ in the world for the *same* retinal image.

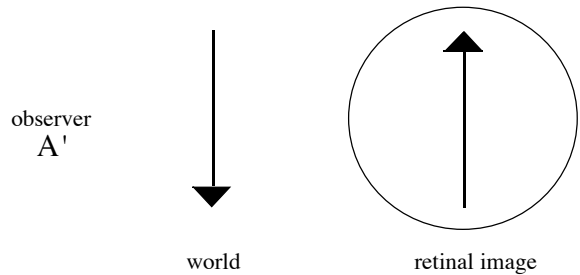

Chapter 8

Although I don't think it's a substantive argument, someone might say, "Well, B is only wearing an optical contraption. Inversion of incoming light only affects B's experience because there's a contraption involved. It really doesn't affect the physical world itself." Well, inversion of incoming light really does affect the physical world because a contraption is just a contraption, a physical instrument involved in the display of light on the observer's retina. It doesn't stop the observer's adaptation to the altered pattern of incoming light on the retina. The orientation of objects and phenomena in the physical world *depends in part on the observer*. That includes, for example, the orientation of something called the spin of an electron along a particular spatial axis. Instead of being built the way we are, we could have been built with an optical apparatus around our heads. Then we all naturally would think of up and down while wearing this contraption on our heads. There is something else I can say if you don't accept my point yet. What if someone's retinas were simply rotated 180º in his eyes? No nerves are severed. Just the retinas are rotated. Then there is no contraption added, and we would still get the same results. Please forgive me for thinking such a thing, but sometimes it takes a strong statement to make a point.

Up can be down and down can be up in the physical world. An electron's spin down along a spatial axis for one observer, for example, can be spin up along this same axis for another observer wearing an apparatus like Stratton's and whose vision has adapted. Whether up is down or down is up depends on how incoming light enters an observer's eyes and whose spatial structure we are considering.

Chapter 9

A Final Word

Physical world, mental world. They are separate and not separate. Both are needed to explain the wonder of our experience of living in the physical world. There are certain important advantages to the results of this study. Einstein (1954) wrote, "the eternal mystery of the world is its comprehensibility" (p. 292). Our comprehension of the world is no longer so mysterious. We don't have this chasm between the physical world and our experience of it. The mind is directly tied to it. Also, you know this chasm that we have supposed exists between the physical world and the mind has contributed to a great sense of meaningless and isolation on the part of people. In the midst of the great cosmos, here we are, apparently insignificant to its functioning. Well, this is not the case. Each of us significantly affects the structure and functioning of the world.

Without us, there would be no reference frame. What would the motion of objects be without a reference frame, whatever physical theory we are considering? The reference frame is essential to the invariance of physical law underlying the conservation laws that have been discussed. Without us, there would be no special relativity, no general relativity, no quantum mechanics, and no statistical mechanics. There would not be an observer at rest in a reference frame, or if you like, a reference frame at rest relative to an observer. There wouldn't be the relativity of simultaneity in special relativity or the principle of equivalence in general relativity. Because we wouldn't have these basic results, we wouldn't have the other results of these theories. I don't even think we would have gravitational mass. We wouldn't have measurements in the physical world corresponding to our different pictures that we saw in quantum mechanics, namely the many skinny hills or the one wide hill. With regard to statistical mechanics, what would probabilistic knowledge mean without someone for whom the probabilities meant something? What would temperature and other characteristics of physical phenomena be that depend on probabilistic knowledge? How would the representative ensemble exist? I didn't mention this to you, but you might have guessed. Those dots that make up the hills we talked about in quantum mechanics are also, to a significant degree, governed by probability. There is also the spatial directionality of the physical world that is dependent on us, a

# A Final Word

result that comes from psychological research.  To sum up, without us, the physical world is difficult to imagine.

Now there is a big challenge before us.  We need to find out more about our relationship to the physical world.  We need to really look at our theories in physics and psychology, and perhaps elsewhere, to see if there are ways that these theories are relevant to domains that they were not developed for.  We have to broaden our viewpoint and not say physics is only about the physical world and psychology is only about the mind and expressions of mind found in bodily action.  We have to accept the fact that the boundaries of physics and psychology broadly overlap.  We have to do experiments that rely on methodology and results from both psychology and physics.  Experiments will answer our questions.  With this perspective, I am confident that significant progress can be made in understanding our relationship to the physical world.

## Sources for the Quotes

## Some Other Sources

# Sources

# Sources

# Sources